\documentclass{ws-p8-50x6-00}

\begin{document}

\title{The Early Universe Odyssey With Gravitational Waves}

\author{L. P. Grishchuk}

\address{Department of Physics and Astronomy, P. O.
Box 913, Cardiff University, CF24 3YB, United Kingdom \\and \\
Sternberg Astronomical Institute, Moscow University, Moscow 119899, Russia\\
E-mail: grishchuk@astro.cf.ac.uk}


\maketitle

\abstracts{
This contribution summarizes some recent work on gravitational-wave astronomy
and, especially, on the generation and detection of relic gravitational waves.
We begin with a brief discussion of astrophysical sources of gravitational
waves that are likely to be detected first by the coming in operation laser
interferometers, such as GEO, LIGO, VIRGO. Then, we proceed to relic
gravitational waves emphasizing
their quantum-mechanical origin and the inevitability of their existence.
Combining the theory with available observations, we discuss the prospects
of direct detection of relic gravitational waves. A considerable part of the
paper is devoted to comparison of relic gravitational waves with the density
perturbations of quantum-mechanical origin. It is shown how the phenomenon of
squeezing of quantum-mechanically generated cosmological perturbations
manifests itself in the periodic structures of the
metric power spectra and in the oscillatory behaviour of the CMBR multipoles
$C_l$ as a function of $l$. The cosmological importance of the theoretically
calculated (statistical) dipole moment $C_1$ is stressed. The paper contains
also some comments on the damage to gravitational-wave research inflicted by the
``standard inflationary result". We conclude with the (now common) remarks
on the great scientific importance of the continuing effort to observe relic
gravitational waves, directly or indirectly.}

\section{Which sources of gravitational waves will be detected first ?}
\label{sec:first}

The comprehensive contributions of Coccia \cite{Coccia} and Giazotto
\cite{Giazotto} have described the outstanding experimental effort that is now
going on at detecting gravitational waves. As we all know, great expectations
are related with the coming in operation laser interferometers: the
British-German GEO600 \cite{geo}, the two American instruments
LIGO \cite{ligo}, and the French-Italian VIRGO \cite{virgo}. It is important
to know in advance, and to be prepared to monitor, those sources
of gravitational waves that are likely to be detected first in the ongoing
and forthcoming observations. Before proceeding to the main topic of my
contribution - relic gravitational waves - I would like
to present a theorist's view on this issue. I will briefly report on the
conclusions of a detailed study undertaken in Ref.\cite{ufn}. By comparing
our theoretical analysis with only 3 out of several experimental programs,
we, of course, do not mean to diminish the importance of work of other
experimental groups.

It is common to divide possible sources of gravitational waves in
three broad categories:
quasi-periodic, explosive, and stochastic. We are aware of definitely
existing astrophysical sources belonging to each category.
To be interesting from the point of view of its detection,
the source should be sufficiently powerful, should fall in the frequency
band of the detector, and should occur reasonably often during the life-time
of the instrument. After having analysed many possible sources, and taking
all the factors into account, the authors of Ref.\cite{ufn} gave
their preference to compact binaries consisting of black holes and
neutron stars. It is clear that in order to radiate large-intensity
gravitational waves at frequencies
accesible to ground-based instruments, the objects forming a pair should
be massive and should orbit each other at very small separations - a few
hundreds kilometers. According to the existing views, such massive objects
can only be the end-products of stellar evolution - neutron stars (NS) and
black holes (BH). Compact binary systems of our interest are
coalescing pairs that are only tens
of minutes away from the final merger, i.e., from the formation of a resulting
black hole or, possibly, from another spectacular event, such as a
gamma-ray burst. The important question is how many such close binaries
exist in our Galaxy and at cosmological distances. This determines the
event rate, that is, the number of coalescence events that are likely
to occur in a given cosmological volume during, say, a 1 year of
observations. The event rate is
partially constrained by the existing observations of binary pulsars, but
its evaluation mostly relies on the numerical modelling of diverse
evolutionary tracks of massive binary stars.

\begin{figure}[t]
\begin{center}
\epsfxsize=20pc 
\epsfbox{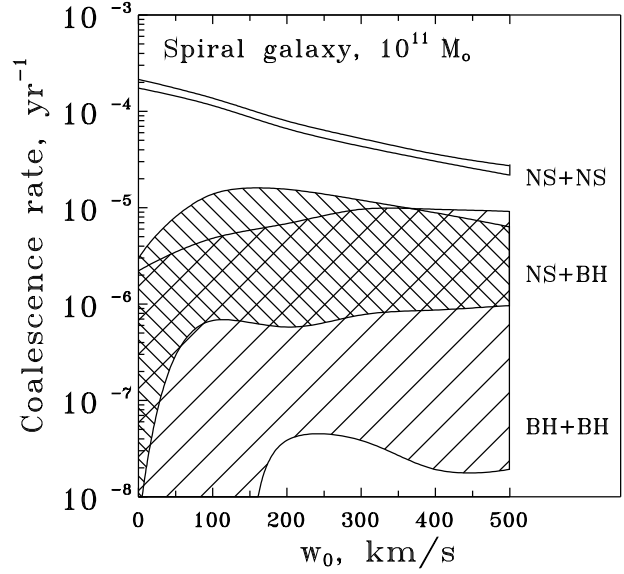} 
\caption{NS+NS, BH+NS, and BH+BH merging rates in a $10^{11} M_\odot$ galaxy as
functions of the kick velocity parameter $w_0$ for the Lyne-Lorimer kick
velocity distribution.}
\label{Fig:od_rate.eps}
\end{center}
\end{figure}

The event rates are sensitive to several evolutionary parameters, one
of which is the so-called kick velocity parameter $w_0$. This parameter
enters the distribution function for velocities that may be imparted
on a neutron star or a black hole
at the moment of its formation. It is believed that $w_0$ lies in the
interval (200 - 400)$km/sec$. In Fig. \ref{Fig:od_rate.eps} we plot the NS+NS,
NS+BH, and BH+BH merging rates as functions of $w_0$. This calculation
has been performed for a typical spiral galaxy of the mass $10^{11} M_\odot$
and under a reasonable choice of other evolutionary parameters \cite{LPP}.

It follows from this graph that the NS+NS rate is expected to be at the
level $3 \times 10^{-5}
year^{-1}$, whereas the BH+BH rate is, at least, one order of magnitude
lower, i.e., $3 \times 10^{-6} year^{-1}$. These rates for a typical
galaxy, ${\cal R}_G$, define the rates for a given cosmological volume,
${\cal R}_V$, which includes many galaxies. When deriving ${\cal R}_V$,
we have used a conservative estimate for the baryon content of the Universe,
and arrived at the relationship
\[
{\cal R}_V = 0.1 {\cal R}_G \left(r \over 1~Mpc \right)^3.
\]
Thus, within the volume of radius $r = 100~Mpc$, and during
1 year, one expects that there will be 3 of the NS+NS coalescences and
only 0.3 of the NS+BH or BH+BH coalescences. The increase of the radius $r$ to
$r=200~Mpc$ increases the volume and the event rates by the factor 8.

The calculated event rates and gravitational wave amplitudes emitted by
individual binaries should be compared with the instrumental sensitivities.
The important fact is that the mass of a typical neutron star is
1.4 $M_\odot$, whereas the mass of a typical stellar black hole is
(10 - 15) $M_\odot$. Thus, a pair of black holes is a more powerful source
of gravitational waves than a pair of neutron stars. The black hole binaries
are less numerous than the neutron star binaries, but they can be seen,
by a given instrument, from much larger distances. The observationally
required situation is when the detector's sensitivity allows one to see
a few events per year, and at the signal to noise ratio $S/N$ level
of, at least, 2. Thus, the problem now is to find out the $S/N$ for
binaries of a given total mass $M$.

\begin{figure}[t]
\begin{center}
\epsfxsize=20pc 
\epsfbox{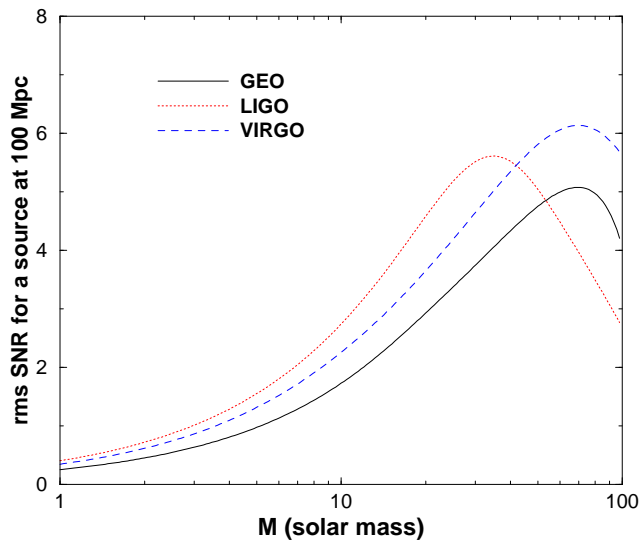} 
\caption{Signal-to-noise ratio, in initial interferometers, as a function
of total mass, for inspiral signals from binaries of equal masses at 100 Mpc
and averaged over source inclination.}
\label{Fig:od_snrnew.eps}
\end{center}
\end{figure}

In Fig. \ref{Fig:od_snrnew.eps} we plot \cite{ufn} the $(S/N)$ as a function
of $M$. We have used the sensitivity curves of the initial interferometers in
LIGO, VIRGO and GEO.  It is assumed that the binaries consist of equal mass
components, so the total mass of a NS+NS system is around 2.8$M_\odot$, while
the total mass of a BH+BH system is around (20 - 30)$M_\odot$. The $S/N$ curves
are shown for binaries placed at $r = 100~Mpc$. The increase of the distance to
$r = 200~Mpc$ decreases the values of the plotted functions by the factor 2.

It is seen from this graph that the $S/N$ is less then 1 for the NS+NS
binaries. Although the required 3 mergers of neutron stars do occur in
the volume with the radius $r=100~Mpc$, these sources cannot be regarded
as detectable by the initial interferometers. On the other hand, for a pair of
black holes with the total mass $M = (20 - 30)M_\odot$, the $S/N$ can be
at the level of 4, or even larger (in the LIGO detectors). The number of
the BH+BH events in the discussed volume is too small, though.
However, the $S/N$ is still at the level of 2, or better than 2,
if the binary black hole is placed at the twice as large distance,
i.e. $r = 200~Mpc$. As was shown above, one can
expect about 2 - 3 BH+BH events in this larger volume. This is why we 
believe~\cite{ufn} that the coalescing black holes will probably be the first
sources detected by the initial laser interferometers.
It appears that other possible
sources of gravitational waves are not likely to be detected at all by these
instruments. The situation will become much better, and for all
known sources, when the advanced interferometers become operational.
And, of course, there exists a nonzero probability that the first detected
signals will be totally unexpected to everyone, or something what the authors
of Ref. \cite{ufn} did not even include in their list of options.
For additional information on
astrophysical sources of gravitational waves and their detection, see the
comprehensive contributions by Ferrari \cite{Ferrari} and Vallisneri
\cite{Vallis}.

Relic gravitational waves are not expected to be accessible to the initial laser
interferometers. The prospects are much better, though, for the advanced
ground-based instruments and for the space-based interferometer LISA \cite{lisa}.
At very large scales, relic gravitational waves should also manifest themselves
through the anisotropy and polarisation of the cosmic
microwave background radiation (CMBR). It is fundamentally important to
discover and study relic gravitational waves. They are the only clue to
the processes that have taken place in the extreme conditions of the
very early Universe and, indeed, they are the only clue to the origin of
the Universe itself. Without
observing  and studying relic gravitational waves, there exists
hardly any other way to test fundamental physical theories, such as
superstrings and $M$-theory. The understanding that the expansion rate of
the very early Universe could be grossly different from the one governed by the
radiation-dominated fluid exists for a long time, at least, from the study
of Sakharov \cite{sakh}. In the recent years, however, there exists the unjust
tendency to replace the notion and all the complexities of the very early
Universe by a single word - ``inflation''. The reminder of my contribution
is devoted to relic gravitational waves and the associated issues of
cosmological perturbations of quantum-mechanical origin.

\section{The inevitability of relic gravitational waves}
\label{sec:inevit}

The theory of almost all astrophysical sources of gravitational waves is
based on classical physics. In contrast, the theory of relic gravitational
waves involves the elements of quantum physics. It is the power and beauty
of relic gravitational waves that their existence relies only on general
relativity and basic principles of quantum field theory, such as the
concept of zero-point quantum oscillations. To understand the generation
of relic gravitons it is convenient to start from a classical picture.
Imagine a classical gravitational wave propagating in a nonstationary
universe. The nonlinear character of the Einstein equations provides
the coupling of the gravitational wave field to the smooth gravitational field
of the evolving universe. The gravitational wave itself can be arbitrarily weak,
so that the quadratic terms of the wave field can be neglected. However,
the nonlinearity of the Einstein equations leads to the presence of the
time-dependent background gravitational field in the coefficients of the
wave equation. The background gravitational field plays the role of the
pump field capable of increasing the amplitude of the propagating
gravitational wave \cite{LPG1}. In fact, the traveling
wave is being amplified, with the accompanying appearance of the wave
traveling in the opposite direction. Together, they form an (almost)
standing wave. This type of coupling of a wave-field to gravity is not
something that is true for any wave-field. On the contrary, it is
quite unusual. For instance, it does not take place for electromagnetic waves.

The quantum element enters the picture when we realise that if the initial
gravitational waves represent only the inevitable zero-point quantum
oscillations, the amplification mechanism should be at work. In
precise technical terms, the quantum-mechanical Schr\"{o}dinger evolution
transforms the initial vacuum state of gravitational waves into a
multi-quantum state known as squeezed vacuum quantum state \cite{GS}.
It is the variance of phase that gets strongly diminished (squeezed) while
the mean number of quanta and its variance get strongly increased.
Of course, for this effect to be numerically considerable,
the coupling of the wave field to the pump field should be considerable.
The coupling is regulated by the ratio of the wave period to the
characteristic time-scale of variations of the background gravitational
field. If the pump field varies very slowly (adiabatically), this
ratio is very small,
and practically nothing happens to the wave. In other words, during the
adiabatic regime, the vacuum state remains the vacuum state. In the
opposite (superadiabatic) regime, the relative increase (as compared
with adiabatic behaviour) of the wave amplitude,
and the increase of the mean number of particles, is most dramatic. In
the cosmological setting, the amplification mechanism is effective
during the interval of time when the wavelength is comparable with, and
longer than, the cosmological Hubble radius. Thus, the generating
process (superadiabatic (parametric) amplification of the
zero-point quantum oscillations \cite{LPG1}) is a local and dynamical
process, it does not rely on the existence of globally defined event
horizons of the background space-time.
Moreover, in cosmological models of practical interest, there is no
horizons at all. It is also irrelevant which particular matter source drives
the gravitational pump field, be it a scalar field or something else.
In some of its basic aspects, the
quantum-mechanical generation of cosmological perturbations is similar
to the laboratory-tested generation of squeezed light.

The gravitational field of a homogeneous isotropic universe is
given by the metric
\begin{equation}\label{cosmet}
{\rm d}s^2 = - c^2{\rm d}t^2 + a^2(t) g_{ij} {\rm d}x^i{\rm d}x^j =
a^2({\eta})[-{\rm d}\eta^2 + g_{ij} {\rm d}x^i{\rm d}x^j] \ \ ,
\end{equation}
where the scale factor $a(t)$ (or $a(\eta)$) is driven by a
matter distribution with some effective (in general, time-dependent)
equation of state. The perturbed gravitational field can be written (for
simplicity, in a spatially-flat universe) as
\begin{equation}
\label{cosmetgw}
{\rm d}s^2 = a^2({\eta})[-{\rm d}\eta^2 + (\delta_{ij} + h_{ij})
{\rm d}x^i{\rm d}x^j]  ,
\end{equation}

\begin{eqnarray}
\label{hij}
h_{ij} (\eta ,{\bf x})
= {{\cal C}\over (2\pi )^{3/2}} \int_{-\infty}^\infty d^3{\bf n}
  \sum_{s=1, 2}~{\stackrel{s}{p}}_{ij} ({\bf n})
   {1\over \sqrt{2n}}
\left[ {\stackrel{s}{h}}_n (\eta ) e^{i{\bf n}\cdot {\bf x}}~
                 {\stackrel{s}{c}}_{\bf n}
                +{\stackrel{s}{h}}_n^{\ast}(\eta) e^{-i{\bf n}\cdot {\bf x}}~
                 {\stackrel{s}{c}}_{\bf n}^{\dag}  \right].
\end{eqnarray}

By expanding the functions $h_{ij} (\eta ,{\bf x})$ over spatial
Fourier harmonics $e^{\pm i{\bf nx}}$ we reduce the perturbed dynamical
problem to the evolution of mode functions
${\stackrel{s}{h}}_n (\eta )$ for each mode ${\bf n}$.
Two polarisation tensors ${\stackrel{s}{p}}_{ij}({\bf n}),~s=1,2$ have
different forms depending on whether they represent gravitational waves,
rotational perturbations, or density perturbations. In the case of
gravitational waves, the polarisation tensors describe two polarisation
states which are often called the ``plus'' and ``cross'' polarisations.

For the quantized field, the quantities
${\stackrel{s}{c}}_{\bf n},~{\stackrel{s}{c}}_{\bf n}^{\dag}$ are
annihilation and creation operators satisfying the conditions
\be\label{comm}
[{\stackrel{s'}{c}}_{\bf n},~{\stackrel{s}{c}}_{{\bf m}}^{\dag}]=
\delta_{s's}\delta^3({\bf n}-{\bf m})\>, \quad
{\stackrel{s}{c}}_{\bf n}|0\rangle =0 \ \ ,
\ee
where $|0\rangle$ (for each ${\bf n}$ and $s$) is the fixed
initial vacuum state defined at some $\eta_0$ in the very distant past, long
before the superadiabatic regime for the given mode has started. In that
early era, the mode functions ${\stackrel{s}{h}}_n (\eta )$
behaved as $\propto e^{-i n \eta}$, so that each mode ${\bf n}$ represented a
strict traveling wave propagating in the direction of ${\bf n}$.
In the case of gravitational waves, the normalization constant ${\cal C}$
is $\sqrt{16 \pi} l_{Pl}$. For cosmological density perturbations, which we
will also discuss below, the normalisation constant is $\sqrt{24 \pi} l_{Pl}$.

The calculation of quantum-mechanical expectation values and correlation
functions provides the link between quantum mechanics and macroscopic
physics.
Using the representation (\ref{hij}) and definitions above, one finds the
variance of metric perturbations:
\begin{equation}
\label{hmvar}
\langle 0| h_{ij}(\eta,{\bf x}) h^{ij}(\eta,{\bf x})|0\rangle
= {{\cal C}^2\over 2\pi^2} \int_{0}^{\infty} n^2\sum_{s=1,2}
|{\stackrel{s}{h}}_n(\eta)|^2 \frac{{\rm d}n}{n}.
\end{equation}
The quantity
\begin{equation}
\label{power}
h^2(n, \eta) =
{{\cal C}^2\over 2\pi^2} n^2\sum_{s=1,2} |{\stackrel{s}{h}}_n(\eta)|^2
\end{equation}
gives the mean-square value of the metric (gravitational field) perturbations in
a logarithmic interval of $n$ and is called the (dimensionless) power spectrum.
The power spectrum of metric perturbations is a
quantity of great observational importance. It defines the temporal structure
and amplitudes of the g.w. signal in the frequency bands of direct experimental
searches. It is also crucial for calculations of anisotropy and
polarisation induced in CMBR by relic gravitational waves and by other
metric perturations. The inevitable
squeezing makes the spectrum an oscillatory function of time, which can also
be thought of as a consequence of the standing-wave pattern of the generated
field. At every fixed moment of time, the spectrum, as a function of the
wave-number $n$, has many maxima and zeros.

To find the power spectrum at any given moment of time (for instance, today
or at the moment of decoupling of CMBR from the rest of matter) we need to
know the mode functions at those moments of time. The mode functions are
governed by the Heisenberg equations of motion whose interaction
Hamiltonian participates with the coupling function proportional to
$a^{\prime}/a$. These equations are, of course, equivalent
to the perturbed Einstein equations. The second-order ``master equation''
describing gravitational waves is \cite{LPG1}
\begin{equation}\label{fieldeq}
{\stackrel{s}\mu}_{n}^{\prime\prime} + {\stackrel{s}\mu}_{n} \left[n^2 -
\frac{a^{\prime\prime}}{a}\right] = 0 \  ,
\end{equation}
where the functions ${\stackrel{s}\mu}_n(\eta)$ are related to the mode
functions ${\stackrel{s}{h}}_n(\eta )$ by
\be \label{mudef}
{\stackrel{s}{\mu}}_n (\eta) \equiv  a(\eta) {\stackrel{s}{h}}_n (\eta ) .
\ee
One can view Eq. (\ref{fieldeq}) as the equation for a parametrically
disturbed oscillator (the term in brackets is the square of its variable
frequency), or as the Schr\"{o}dinger equation for a particle moving in
the presence of a potential barrier $W(\eta)= a^{\prime \prime}/a$ (while
remembering that $\eta$ is a time coordinate rather than a spatial coordinate).

As soon as the pump field (represented by $a(\eta)$) is known, and since the
initial conditions are fully determined, the mode functions are strictly
calculable. We know relatively well the behaviour of $a(\eta)$ at the
matter-diminated $(m)$ and radiation-dominated $(e)$ stages of expansion.
However, we do not know $a(\eta)$ at the most important preceeding stage
of evolution, which we call the initial $(i)$ stage. Ideally, if
the full evolution of $a(\eta)$ had followed from some fundamental
physical theory, the relic g.w. background would be uniquely determined, and we
would have to live with what the fundamental theory dictates.
In the absence of such a theory, we can only do calculations for various
possible models and compare the results with observations. Conversely, the
observational restrictions on relic gravitons, or their direct detection, is
our clue to the dynamics of the very early Universe.

\section{Calculating the power spectrum}
\label{sec:spectrum}

The scale factor $a(\eta)$ at the matter-dominated stage,
governed by whatever matter with the effective equation of state $p=0$,
behaves as $a(\eta) \propto \eta^2$. It is convenient to write $a(\eta)$
in the explicit form
\be\label{scale-matter}
a(\eta) = 2l_H (\eta-\eta_m)^2 \  ,
\ee
where $l_H$ is the Hubble radius today and $\eta_m$ is a constant explained
below. The moment of time ``today" (in cosmological sense) is labeled
by $\eta=\eta_R$ (the subscript $R$ denoting ``reception'').
It is convenient to choose
\begin{equation} \label{etaR}
\eta_R - \eta_m =1.
\end{equation}
With this convention,
$a(\eta_R) = 2 l_H$, and the wave, of any physical nature, whose
wavelength $\lambda$ today is equal to today's
Hubble radius, carries the constant wavenumber $n_H = 4 \pi$. Longer waves
have smaller $n$'s and shorter waves have larger $n$'s,
according to the relationship $n = 4 \pi l_H/ \lambda$. For example,
the ground-based gravitational wave detectors are most sensitive to
frequencies around $30 Hz - 3000 Hz$. The corresponding wavelengths have
wavenumbers $n$ somewhere in the interval $10^{20} - 10^{22}$.

The matter-dominated era
was preceded by the radiation-dominated era with the scale factor
$a(\eta) \propto \eta$. Without any
essential loss of generality, we assume that the transition from $e$ era
to $m$ era was instantaneous and took place at some $\eta = \eta_2$. The
redshift of the transition is
$z_{eq}$: $a(\eta_R)/a(\eta_2) = 1 + z_{eq}$. It is
believed that $z_{eq}$ is somewhere near $6 \times 10^3$. In its turn, the
radiation-dominated era was preceded by the initial era of expansion,
whose nature and
scale factor are, strictly speaking, unknown. To simplify the analysis,
and since the wave equations admit simple exact solutions in the case of
power-law scale factors, we assume that the $i$ era, similar to the
$e$ and $m$ eras, was also described by a power-law scale factor.
Following the early notations \cite{LPG1}, we parametrize the $i$ era
by $a(\eta) \propto |\eta|^{1+\beta}$. The
transition from $i$ era to $e$ era takes place at some $\eta = \eta_1$
and at redshift $z_i$: $a(\eta_R)/a(\eta_1) = 1+ z_{i}$. Further analysis
shows (see below) that in order to get the right amplitude of the generated
perturbations, the numerical value of $z_i$ should be
somewhere near $10^{30}$. Whether the $i$ era was governed by a scalar
field $\varphi$ (the central element of inflationary scenaria) or by
something else, is irrelevant. What is relevant is the functional form of
the gravitational pump field represented by the scale factor $a(\eta)$.

We now write the full evolution of the growing scale factor explicitly:
\be
   a(\eta )= l_0 | \eta |^{1+\beta}, \quad
   \eta \leq \eta_1,~~\eta_1 <0,~~\beta < -1 \ \ ,
\ee
\be
    a(\eta )= l_0 a_e (\eta - \eta_e) , \quad
    \eta_1 \leq \eta \leq \eta_2 \ \ ,
\ee
\be
   a(\eta )=  2l_H (\eta - \eta_m )^2 , \quad \eta_2 \leq \eta .
\ee
The continuous matching of $a(\eta)$ and $a^{\prime} (\eta)$ at the
transition points determines all the participating constants
in terms of $l_H,~z_i,~z_{eq}$ and $\beta$. Thus, we are essentially left
with the two unknown parametrs: $z_i$ and $\beta$. The $z_i$ is primarily
responsible for the amplitude of the generated perturbations, while the
$\beta$ is responsible for the spectral slope of the metric power spectrum.
If $\beta = -2$, the generated spectrum is flat, that is, independent of
the wave-number $n$. The flat spectrum is also called the
Harrison-Zeldovich-Peebles spectrum and is characterised by the
spectral index ${\rm n} =1$ .
The relationship between the spectral index ${\rm n}$ and parameter
$\beta$ is ${\rm n} = 2\beta +5$. Inflationary models governed by scalar
fields  $\varphi$ are uncapable of producing the ``blue" power spectra, i.e.
spectra with ${\rm n} > 1$ ($\beta > -2$). Indeed, $\beta > -2$ requires the
effective equation of state at the $i$ stage to be $\epsilon + p < 0$,
but this cannot be accomodated by any scalar field whatever the
scalar field potential $V(\varphi)$ may be \cite{ufn}.

For the CMBR calculations one also needs the redshift $z_{dec}$ of the
last scattering surface $\eta = \eta_E$ (with the subscript $E$ denoting
``emission''), where the CMBR photons have
decoupled from rest of the matter: $a(\eta_R)/a(\eta_E) = 1+ z_{dec}$.
The numerical  value of $z_{dec}$ is somewhere near 1000.

Since the scale factor and initial conditions for the mode functions
have been strictly defined, the power spectrum (\ref{power}) is
unambiguously calculable. We will present the results of calculation of
$h^2(n, \eta)$ taken at the time of decoupling $\eta = \eta_E$.
The exact form of the power spectrum is
\be \label{powexact}
h^2 (n, \eta_E) = {l_{Pl}^2 \over 4 \pi l_H^2}(1+z_{dec}) n^4 |B|^2 g^2 (x, b)
\ \ ,
\ee
where
\[
g^2(x,b) = \left[\rho_{1}(x) j_{1}(bx) - \rho_{2}(x) j_{-2} (bx)\right]^2,~~
x \equiv {n \over 2\sqrt{1+z_{eq}}}, ~~
b \equiv {2 \sqrt{1+z_{eq}} \over \sqrt{1+z_{dec}} }\,.
\]
\[
\rho_{1}(x) = \frac {1}{x^2}\left[(8x^2 -1)\sin x + 4x \cos x +\sin 3x \right],
\]
\[
\rho_{2}(x) = \frac {1}{x^2}\left[(8x^2 -1)\cos x- 4x \sin x +\cos 3x \right],
\]
and $j_{1}(bx),~j_{-2} (bx)$ are spherical Bessel functions. The quantity
$n^4|B|^2$ is, in general, $n$-dependent and $\beta$-dependent, 
but for the case of
$\beta = -2$ it simplifies to
\be \label{|B|2}
n^4|B|^2 = \frac {4(1+z_i)^4}{(1+z_{eq})^2},~~~~~\beta =-2.
\ee

\begin{figure}[t]
\begin{center}
\epsfxsize=20pc 
\epsfbox{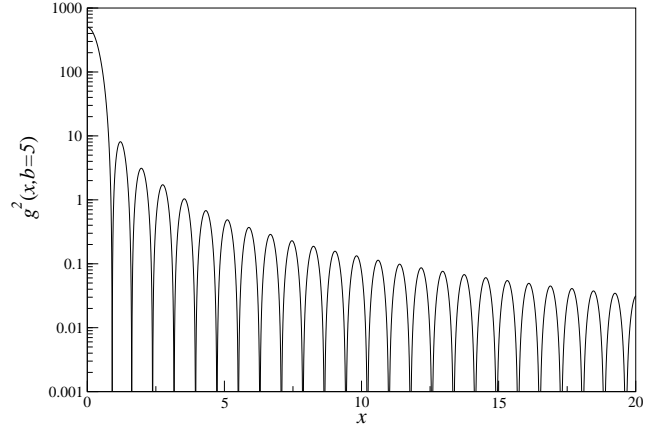} 
\caption{Plot of $g^2(x,b=5)$ versus $x$.}
\label{fig:od_gSquare.eps}
\end{center}
\end{figure}

In Fig. \ref{fig:od_gSquare.eps} we show \cite{bg} the function
$g^2(x,b)$ for $b = 5$.

\section{Detectability of relic gravitational waves}
\label{sec:detect}

We start from a discussion of manifestations of relic gravitational
waves in the CMBR
anisotropies. It can be shown (see \cite{bg} and references there) that
the angular correlation function has the universal form
\begin{equation}\label{Tmom}
   \left\langle 0\left| {\delta T \over T}
   ({\bf e}_1) {\delta T \over T} ({\bf e}_2)\right|0 \right\rangle
 = \sum_{l=2}^\infty \frac{2l+1}{4\pi} C_l P_l(\cos\delta ),
\end{equation}
where $P_l(\cos\delta )$ are Legendre polynomials for the separation angle
$\delta$ between the unit vectors ${\bf e}_1$ and  ${\bf e}_2$, and
the multipole moments $C_l$ are explicitely determined by the metric
mode functions.

\begin{figure}[t]
\begin{center}
\epsfxsize=20pc 
\epsfbox{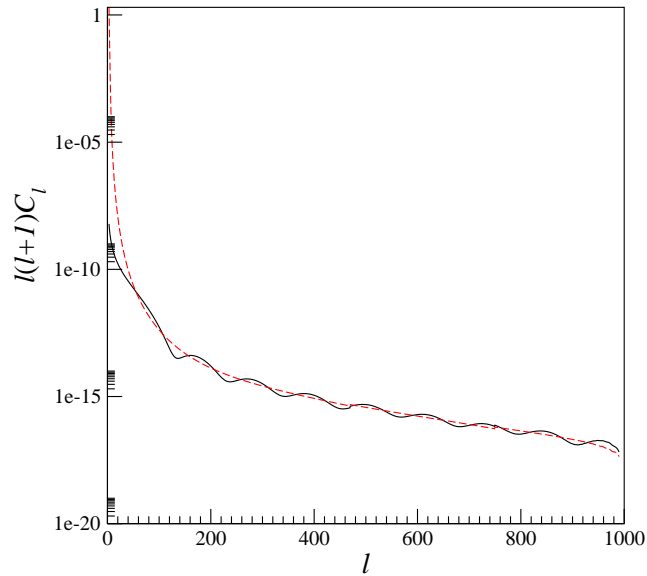} 
\caption{The solid line depicts the plot of $l(l+1)C_l$ versus $l$,
normalized such that at $l=10$, we have $l(l+1)C_l = 6.4\times 10^{-10}$, which
tallies with observations. The dashed line is the corresponding plot
for the non-squeezed g.w. background. We take $\beta = -2$, and the redshifts
$z_{eq} = 10^4$ and $z_{dec} = 10^3$.}
\label{fig:od_Cl.eps}
\end{center}
\end{figure}

In Fig. \ref{fig:od_Cl.eps} we show \cite{bg} by a solid line the
graph of the function $l(l+1) C_l$ calculated for the g.w. background with
the power spectrum (\ref{powexact}).
The cosmological parameters were taken, for illustration,
as $z_{eq} = 10^4,~ z_{dec}= 10^3,~ \beta = -2$. The parameter $z_i$ is 
adjusted in such a manner ($z_i = 10^{29.5}$) that the graph goes 
through the point $l(l+1)C_l = 6.4 \times 10^{-10}$ at $l=10$, which 
agrees with observations.
For comparison, the dashed line shows the same function, but calculated
for the alternative (non-squeezed) background of gravitational waves of the
same averaged power density. The remarkable (even if expected) result is that
the background of non-squeezed (traveling) gravitational waves does not produce
oscillations in the angular power spectrum $C_l$, whereas the
background of squeezed (standing) gravitational waves does. The peaks and
dips of the angular power spectrum are in close relationship with the
maxima and zeros of the metric power spectrum of Fig. \ref{fig:od_gSquare.eps}.
It is unlikely that the peaks and dips caused by relic
gravitational waves can be observationally revealed, but this calculation
serves as a guidance for the related explanation (see below) of the actually
observed peaks and dips \cite{peaks} that lay at a higher level of the signal.

\begin{figure}[t]
\begin{center}
\epsfxsize=20pc 
\epsfbox{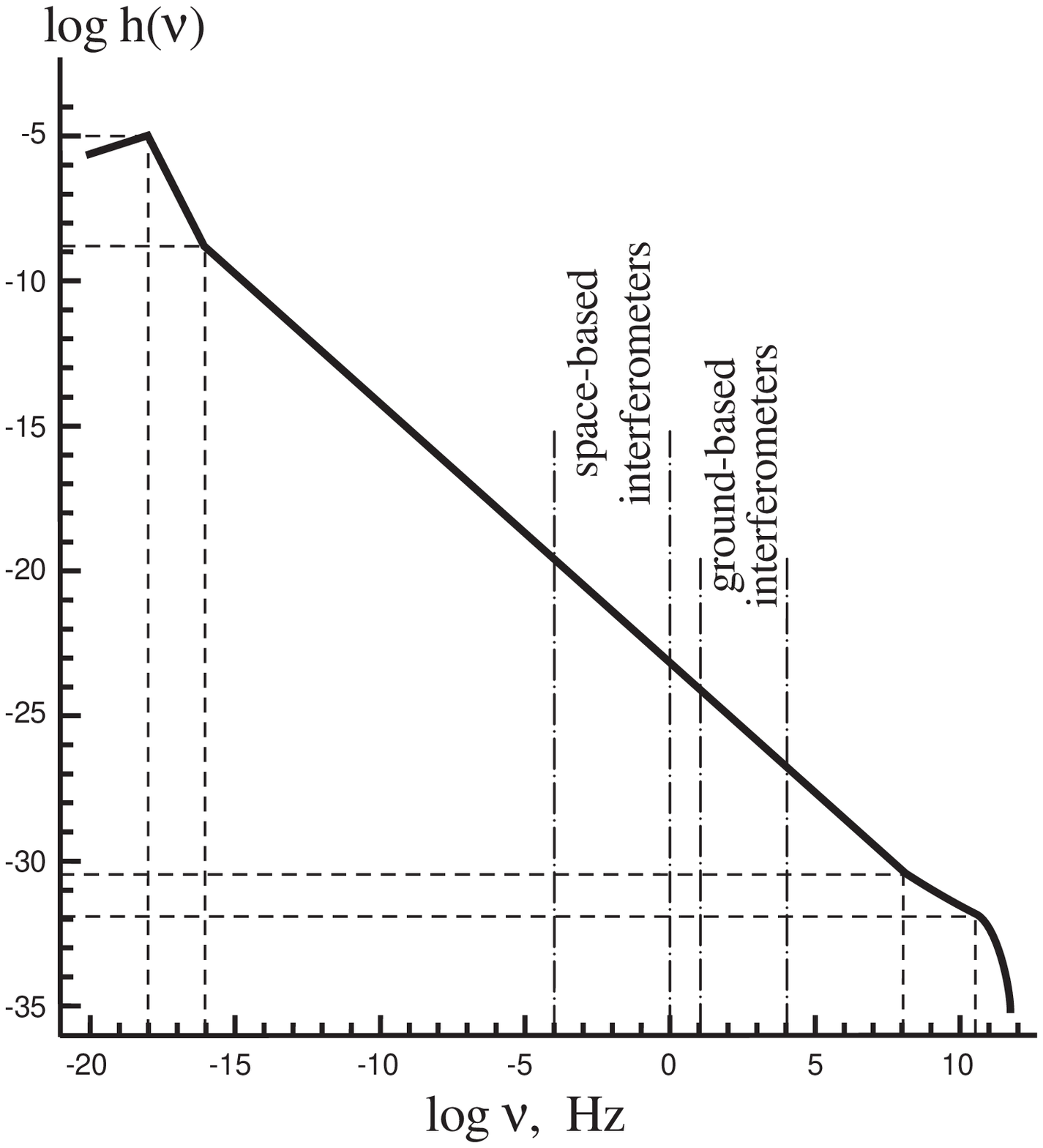} 
\caption{Expected spectrum $h(\nu)$ for the case $\beta = -1.9$.}
\label{fig:od_hnu.eps}
\end{center}
\end{figure}

For purposes of direct experimental searches for relic gravitational
waves, one needs
to know the metric power spectrum in the present era.
In Fig. \ref{fig:od_hnu.eps} we
show the piece-wise envelope of the r.m.s. quantity
$h(n, \eta_R)$ plotted as a function of frequency $\nu$ in $Hertz$.

This particular graph is derived under the assumption that the
parameter $\beta$ is $\beta = -1.9$, which
corresponds to the spectral index ${\rm n} = 1.2$. A few years ago
(see, for example, \cite{bennet}), this value of ${\rm n}$ was
quoted as the one that follows from the lower-order multipoles of the
CMBR data. The most recent analyses, which include data on the large scale
structure, seem to prefer a smaller value of ${\rm n}$. Although the often
cited central values of ${\rm n}$ are still somewhat larger than 1,
the 1$\sigma$ error-bars do not usually include the value ${\rm n} = 1.2$.
One should remember, however, that the current analyses of the data are
consistent only if one assumes that about 70 percent
of ``matter" in the present Universe resides in the cosmological
$\Lambda$ term or in a ``dark energy/quintessence",
which is not subject to gravitational clustering.
It is still not clear whether this conclusion is one of the greatest
discoveries about the Universe, or it may be a result of improper
evaluation of the role of metric perturbations in the CMBR analysis.
We will comment on this issue in the next section. In addition, one should
bear in mind that the flat (${\rm n} =1$) and ``red" (${\rm n} < 1$)
spectra possess a theoretical difficulty: the integral over
wavenumbers $n$ for the mean square value of the metric (gravitational field)
perturbations diverges at the lower limit $n \rightarrow 0$, i.e. for very
long waves. The removal of these divergencies would require some extra
assumptions. In any case, we will discuss the prospects of direct
detection of relic gravitational waves assuming that the graph in
Fig. \ref{fig:od_hnu.eps} is correct or close to correct.

According to analytical formulas and Fig. \ref{fig:od_hnu.eps}, the
theoretically expected r.m.s. value at $\nu = 100 Hz$ is $h_{th} = 10^{-25}$.
The target value of the LIGO-II \cite{ligo2} at this frequency is
$h_{ex} = 10^{-23}$. The remaining gap of 2 orders of magnitude can be
covered by the cross-correlation of sufficiently long strips of data from
two or more detectors. The $S/N$ ratio will be better than 1, if the
common integration time exceeds $\tau = 10^6 sec$. This does not look like
a hopeless task. The situation is somewhat better in the LISA frequency
interval. According to evaluations of Ref. \cite{ufn}, a 1-year
observation time can provide a $S/N$ ratio around 3 in the frequency
interval $(2 \times 10^{-3} - 10^{-2}) Hz$. These estimates do not include
a possible improvement in the $S/N$, which can be achieved if one succeeds
in finding a sophisticated method of exploiting, at these frequencies, the
ever-present squeezing. This improvement would certainly
be possible in a very narrow-band detector, but it is unclear whether
it can ever be realised in a realistic broad-band detector
(for a discussion, see \cite{afp} and \cite{ufn}). At the present time,
it is difficult to say at which scales the relic gravitational waves will
be first discovered. This can happen in direct experiments at relatively
small scales, or, indirectly, at cosmological scales, through the
anisotropy and polarisation measurements of the CMBR.

\section{Relic gravitational waves and primordial density perturbations}
\label{sec:density}

The superadiabatic (parametric) mechanism is always operative for gravitational
waves, but its applicabilty to density perturbations requires special
assumptions about properties of matter. Of course, the
notion of density perturbations includes the associated metric
perturbations $h_{ij}$ of Eq. \ref{hij}; the matter and gravitational field
(metric) variables are always linked by the perturbed Einstein equations.
If the $i$ stage was governed
by a scalar field $\varphi$ coupled to gravity in a special manner
(a hypothesis usually
assumed in inflationary scenaria), the quantum-mechanical generation of
primordial density perturbations, in addition to the inevitable generation
of relic gravitational waves, becomes possible. The underlying physical
theory for these two respective fields is exactly the same, and therefore
the general properties of relic gravitational waves and
primordial density perturbations (squeezing, standing-wave
pattern, spectral slopes, numerical values of the amplitudes, etc.)
are almost identical. In particular, relic gravitational waves and
primordial density perturbations should provide approximately equal
contributions to the lower-order multipoles (starting with
the quadrupole moment $C_2$) of the CMBR anisotropies \cite{gr94}. Some
differences arise, however, at the late stage of
their evolution. In the matter-dominated era, and for wavelengths comfortably
shorter than the Hubble radius, numerical values of the metric components
associated with density perturbations grow in proportion to $a(\eta)$
(gravitational instability), whereas the amplitudes of gravitational waves
decrease in proportion to $1/a(\eta)$ (adiabatic behaviour).
This explains why the metric amplitudes of these two fields, being of the same
order of magnitude at scales comparable with the prersent-day Hubble radius,
are significantly different at much smaller scales. This difference makes
gravitational waves subdominant in their contribution to the CMBR multipoles
near the peak at $l \sim 200$, and makes them so small and difficult to
detect at laboratory scales.

As a benchmark for the
spectral index one can take $\beta = -2$, i.e. flat spectrum.
The CMBR anisotropies caused by gravitational waves are shown
in Fig. \ref{fig:od_Cl.eps}. The lowest excited multipole is $l=2$ (quadrupole).
In contrast, density perturbations produce all multipoles,
including $l=0$ (monopole) and $l=1$ (dipole). In general, CMBR anisotropies
caused by density perturbations have contributions from 3 different
ingredients (for a recent detailed discussion,
see \cite{wein}). Specifically, the ingredients are: the accompanying metric
perturbations (mostly driven by the gravitationally dominant matter, which
is, according to the existing views, a presureless cold dark matter (CDM)),
the intrinsic temperature variations at the last scattering surface,
and the velocities (with respect to the coordianate system
comoving with the CDM) of the last scattering electrons.

The angular power spectrum $l(l+1)C_l$ is usually presented as an outcome
of numerical calculations, where all contributions are being mixed up.
However, the evolutionary equations, on which
the numerical codes are based, allow analytical calculations for the
lower multipoles, i.e. for $l$'s up to $l \approx (30 - 40)$. It is
often claimed that the analytical limit of the employed evolutionary
equations produces the lower-order multipoles $C_l$ according to the formula
\begin{equation}
\label{cl}
C_l = \frac{const.}{l(l+1)}.
\end{equation}
This formula predicts an infinitely large monopole moment $C_0$, and the
dipole moment $C_1$ only a factor 3 greater than the quadrupole moment
$C_2$; both predictions are in a severe conflict with observations.
Moreover, this formula misses the quite steep growth, beginning from
$l=2$, of the function $l(l+1)C_l$, which must take place for purely
gravitational reasons, that is, when the metric perturbations are
treated correctly \cite{dg}.

\begin{figure}[t]
\begin{center}
\epsfxsize=20pc 
\epsfbox{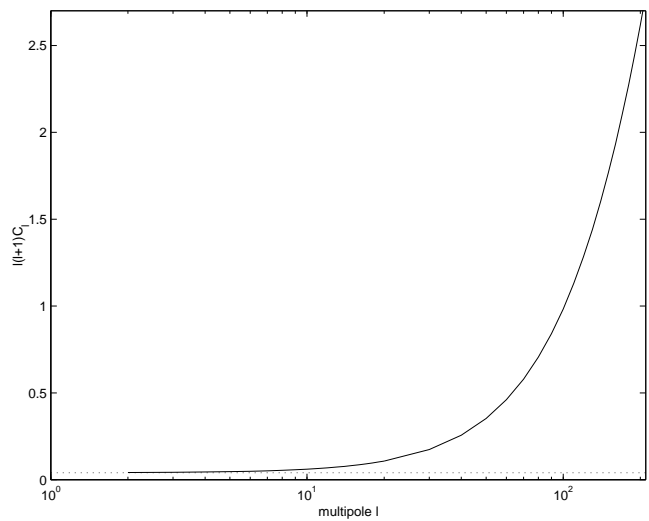} 
\caption{The multipole distributions as described by the exact formula
(solid line) and formula \ref{cl} (dotted line) for the flat primordial
spectrum $\beta = -2$ (${\rm n} = 1$).}
\label{fig:od_andr1.eps}
\end{center}
\end{figure}

It was demonstrated \cite{dg} that formula \ref{cl} is incorrect and
cannot be the outcome of correct equations. Specifically, formula \ref{cl}
arises only if one unjustifiably neglects 3 out of 4 terms in the exact
formula (43) of the Sachs-Wolfe paper \cite{sw}. The correct handling
of metric perturbations makes the monopole moment $C_0$ finite and small,
whereas the dipole moment $C_1$ becomes 5 orders of magnitude greater than the
number following from formula \ref{cl}, and, thus, the theoretical
(statistical) $C_1$ comes in agreement with the observed value of $C_1$.
It was also shown that the growth of the function $l(l+1)C_l$ would be
unlimited, if one were allowed to extrapolate the flat spectrum to arbitrarily
large wave-numbers $n$. This is illustrated in Fig \ref{fig:od_andr1.eps}.

In reality, however, the modulating (transfer) function of the metric
perturbations bends the angular spectrum down and introduces a series of
peaks and dips, analogous to the peaks and dips caused by
gravitational waves in Fig. \ref{fig:od_Cl.eps}.
We will discuss peaks and dips shortly, but first we shall say a few words
about the importance of theoretical calculations of the expected
(statistical) value of the dipole moment $C_1$. Because of statistical
properties of the generated perturbations, all statistical
multipoles $C_l$ are independent of the observer's position.

The derivation of formula \ref{cl} takes into account the perturbations of all
wavelengths. The integral over $n$, which gives rise to this formula, is
formally extended from $n = 0$ to arbitrarily large $n$'s. If this
formula were correct, it would be applicable all over the linear regime of
cosmological perturbations and could possibly break down
only due to nonlinearities at relatively short scales. The gap in 5 orders
of magnitude between the statistical prediction of this formula and the actually
observed $C_1$ would need to be attributed either to extremely improbable
realisation of the perturbed gravitational field, or to the quite
local conditions where nonlinearities set it and where the linear
approximation is not valid. In the latter case, numerical
values of $C_1$'s observed by inhabitants of the not-too-distant galaxies
would not need to have anything in common with each other and with
the $C_1$ observed by us. In contrast, calculations
of Ref. \cite{dg} show that numerical value of the statistical $C_1$ is
most sensitive to scales where the metric power spectrum bends down.
In other words, the corresponding integral is primarily accumulated at
scales of the order of $(100 - 200) Mpc$, i.e. well within the linear
regime. This result comes about from the ``gradient of gravitational potential"
terms (see also \cite{bg}), which are totally
neglected in the derivation of formula \ref{cl}. This means that the
galaxies within, at least, the radius of $(100-200) Mpc$, and, at least,
the field galaxies, are expected to have approximately equal observed $C_1$'s.
This number should also be of approximately the same value as our own
$C_1$ (as soon as our measured $C_1$
turns out to be close to the statistical $C_1$). In principle, this
picture is observationally testable. Of course, we cannot place ourselves
at different galaxies and measure the $C_1$ there. But, it is not excluded
that the CMBR dipoles seen at other galaxies can be evaluated indirectly,
possibly with the help of the Sunyaev-Zeldovich effect. It would be
very interesting if these observational
evaluations could be done. In any case, the calculation of the
theoretical (statistical) $C_1$ is as important as the calculation of other
multipoles. It would be embarassing if some of currently popular
cosmological models would predict the statistical $C_1$ way out of what
is actually observed by us and by observers in other
galaxies (assuming that we can evaluate their dipoles).

\begin{figure}[t]
\begin{center}
\epsfxsize=20pc 
\epsfbox{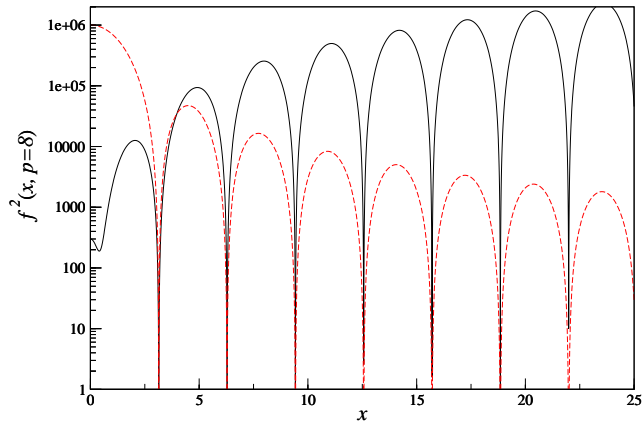} 
\caption{The plot depicted by the solid line is that of $f^2(x,p=8)$
versus $x$. The dashed line shows the behavior of the function
$(\sin x/ x)^2 \times 10^6$.}
\label{fig:od_fSquare.eps}
\end{center}
\end{figure}

It is argued in Ref. \cite{bg} that the (squeezed) metric perturbations
associated with density perturbations are mostly responsible for
the recently observed peaks and dips in the angular power spectrum
$l(l+1)C_l$. This interpretation is different from the concept of ``acoustic
peaks" based on the existence of the plasma sound waves at the last
scattering surface. The rigorous evolution of the quantum-mechanically
generated density perturbations through the radiation-dominated and
matter-dominated eras, allows one to find the metric power spectrum,
defined by Eq. \ref{power}, at the last scattering surface $\eta= \eta_E$.
The analytical formula for $h^2(n, \eta_E)$ is given by
\be \label{powermE}
h^2 (n, \eta_E) = {{\cal C}^2\over 2\pi^2} {n^4|B|^2 (1+z_{eq}) \over
48 l_H^2} \left({\sin y_2 \over y_2}\right)^2 {(300 - 20 p^2 y_2^2 +p^4 y_2^4)
\over 200},
\ee
where
\[
y_2 = \frac{n}{2 \sqrt{3} \sqrt{1+z_{eq}}},~~~ {\rm and}~~~
p  \equiv {2 \sqrt{3} \sqrt{1+z_{eq}} \over \sqrt{1+ z_{dec}} }.
\]
The graph of the oscillating function
\[
f^{2}(x, p) \equiv \left({\sin x \over x}\right)^2 \left[ 300 -20 p^2 x^2 +
p^4 x^4\right],
\]
where $x \equiv y_2$ and, for illustration, we take $p=8$, is shown
in Fig. \ref{fig:od_fSquare.eps}.

The oscillations in the metric power spectrum
are a direct consequence of squeezing (standing-wave pattern) of the
primordial density perturbations. These oscillations find their reflection
in the peaks and dips of the angular power spectrum. The main difference
with the concept of ``acoustic peaks'' lays in the two facts: the amplitudes
of metric perturbations are larger than the amplitudes of matter density
variations in the photon-electron-baryon fluid, and the wave-number periodicity
of metric perturbations is governed by a sine function, instead of a cosine
function for acoustic perurbations. It is argued \cite{bg} that the actually
observed features are better described by the derived analytical formulas
than by the concept of ``acoustic peaks''.

The study of density perturbations is directly relevant to the problem of relic
gravitational waves. If cosmological perturbations are indeed generated
quantum-mechanically (what the present author believes is likely to be true),
many properties of primordial density perturbations and relic gravitational
waves should be common. By studying density perturbations (for instance,
by proving the presence of squeezing) we will be more confident about the
properties of relic gravitational waves.

One should be aware that the story of quantum-mechanically generated
cosmological perturbations is dramatically different in inflationary scenario.
Inflationists love to speak about ``stretching the quantum fluctuations of 
the inflaton to cosmological scales''. As a result of this ``stretching'',
they manage to produce the arbitrarily large amplitudes of today's density
perturbations. Indeed, the ``standard result'' of inflationary scenario
predicts the infinitely large density perturbations, in the limit of the
flat spectrum (${\rm n}=1$), through the set of evaluations:
$\delta \rho / \rho \sim h_S \sim \zeta \sim
H^2/\dot \varphi \sim V^{3/2}(\varphi)/ V'(\varphi) \sim 1/\sqrt{1 -{\rm n}}$.
[The ``quantization'', as understood by inflationists, of the necessarily
coupled system of scalar field and metric perturbations is such, that
the so-called Bardeen's gauge-invariant quantity $\zeta$ is
arbitrarily large already at the beginning of the amplifying
(superadiabatic) regime, and then, being ``conserved'',
this arbitrarily large number translates to the end of the amplifying regime,
and to the present time.] Of course, this ``standard result''
is in full disagreement not only with the theoretical
quantum mechanics, but with available observations too, as long as
the error-boxes of the observationally derived spectral index ${\rm n}$
are centered at ${\rm n} \approx 1$ and include
the value ${\rm n}=1$. However, by composing the ratio of
the gravitational wave amplitude $h_T$ to
the predicted divergent amplitude of the scalar metric perturbations $h_S$
(the so called ``consistency relation'': $h_T/h_S \approx \sqrt{1- {\rm n}}$),
inflationary theorists substitute their prediction of arbitrarily large
density perturbations for the claim that it is the amount of gravitational
waves that should be zero, or almost zero, at cosmological scales and,
hence, down to laboratory scales. This claim has led to many years of
mistreatment of a possible g.w. contribution to the CMBR data. The
``standard inflationary result'' is maintained by inflationists and
their followers until now. For instance, the recent review paper \cite{hudo}
claims that the initial spectrum of gravitational waves is ``constrained
to be small compared with the initial density spectrum''. It
is only in a few recent papers (for example, \cite{efs}) that the
inflationary ``consistency relation'' is not being used when analyzing
the CMBR and large scale structure observations, with some interesting
conclusions. The inflationary claim of the ``negligible'' amount of
gravitational waves influences also the LISA community. It is
sometimes opined that the LISA parameters should be better optimised for
detection of g.w. noise from the multitude of binary white dwarfs,
rather than for a search for relic gravitational waves. There seems
to be some logic behind this opinion. If one hears so often
about ``excellent agreement''  of inflationary predictions with
observations, and if the central of these predictions is ``negligible''
amount of relic gravitational waves, there is no need to bother about them.
Many of the contemporary writers manage to do both: to praise inflationary
predictions and to speculate about ``testing inflation'' with the help
of gravitational waves, including the CMBR polarisation measurements
and post-LISA missions. Apparently, they plan to do this at the signal to
noise ratio level of $10^{-20}$, or so, for ``inflation predicted''
gravitational waves. For some reason, those authors do not suggest to test
inflationary predictions right from the predicted divergent amplitudes
of density perturbations. Sometimes, inflationary calculations based on
the ``standard result'' and ``consistency relation'' arrive at gravitational-wave
amplitudes which are only 1-2 orders of magnitude smaller than, or even
comparable (for the spectral index ${\rm n}$ sufficiently smaller
than 1) with the metric amplitudes of density perturbations. Of course,
this does not make the fundamentally wrong ``standard result'' a correct one.
One can find more reading about the ``standard inflationary result'' and
its critique, including the earlier claim of inflationists that their
divergent amplitudes of density perturbations are justified by the
(incorrect and now abandoned) concept of ``big amplification during
reheating'', in the end of Sec. 6 of Ref. \cite{ufn} and in references there.

\section{Conclusions}
\label{sec:concl}

It is important to understand that the existence of relic gavitational waves
relies only on general relativity and basic principles of quantum field
theory. The detection of relic gravitational waves will be a direct probe
of the expansion rate of the very early Universe, and not a test of
some extra assumptions, such as the dominance of a scalar field or the form
of its potential $V(\varphi)$. If the observed large-scale CMBR anisotropy
$\delta T/ T$ is indeed caused by cosmological perturbations of
quantum-mechanical origin, the contribution of relic gravitational
waves to $\delta T/ T$ must be considerable. The
extrapolation of the g.w. metric power spectrum to shorter wavelengths shows
that, depending on the still existent uncertainty in the spectral index,
relic gravitational waves may be measurable by the advanced ground-based or
space-based laser interferometers. The ever-present squeezing must manifest
itself in the periodic structure of the metric power spectra and, at
cosmological scales, in the oscillatory behaviour of the CMBR multipoles
$C_l$ as a function of $l$. When evaluating the scientific importance
and prospects of discovery of relic gravitational waves, it is necessary
to respect the conclusions of quantum mechanics and general relativity, 
and not the statements of popular fallacies.

\section*{Acknowledgments}
I am grateful to my coauthors whose work was reviewed in this contribution.

\end{document}